\documentclass[aps, prb, superscriptaddress, twocolumn]{revtex4-1}
\usepackage{hyperref}
\usepackage{graphicx}
\usepackage{amsmath}
\usepackage{amssymb}
\usepackage{amsfonts}   
\usepackage{multirow}
\usepackage[b]{esvect} 
\usepackage{braket}
\usepackage{color}
\usepackage{bm}

\newcommand{\bea}{\begin{eqnarray}}
\newcommand{\eea}{\end{eqnarray}}
\newcommand{\beq}{\begin{equation}}
\newcommand{\eeq}{\end{equation}}

\newcommand{\up}{\uparrow}
\newcommand{\down}{\downarrow}

\begin{document}
\title{Influence of electron-phonon coupling on low temperature phases of metallic single-wall carbon nanotubes}

\author{Junichi Okamoto}
\affiliation{Physikalisches Institut, Albert-Ludwigs-Universit\"at Freiburg, Hermann-Herder-Stra\ss e 3, 79104 Freiburg, Germany}

\author{Ludwig Mathey}
\affiliation{Zentrum f\"ur Optische Quantentechnologien and Institut f\"ur Laserphysik, Universit\"at Hamburg, 22761 Hamburg, Germany}
\affiliation{The Hamburg Centre for Ultrafast Imaging, Luruper Chaussee 149, 22761 Hamburg, Germany}

\author{Wen-Min Huang}
\email{wenmin@phys.nchu.edu.tw}
\affiliation{Department of Physics, National Chung-Hsing University, Taichung 40227, Taiwan}

\date{\today}

\begin{abstract}
We investigate the effect of electron-phonon coupling on low temperature phases in metallic single-wall carbon nanotubes. We obtain low-temperature phase diagrams of armchair and zigzag type nanotubes with screened interactions with a weak-coupling renormalization group approach. In the absence of electron-phonon coupling, two types of nanotubes have similar phase diagrams. A $D$-Mott phase or $d$-wave superconductivity appears when the on-site interaction is dominant, while a charge-density wave or an excitonic insulator phase emerges when the nearest neighbor interaction becomes comparable to the on-site interaction. The electron-phonon coupling, treated by a two-cutoff scaling scheme, leads to different behavior in two types of nanotubes. For strong electron-phonon interactions, phonon softening is induced and a Peierls insulator phase appears in armchair nanotubes. We find that this softening of phonons may occur for any intraband scattering phonon mode. On the other hand, the effect of electron-phonon coupling is negligible for zigzag nanotubes. The distinct behavior of armchair and zigzag nanotubes against lattice distortion is explained by analysis of the renormalization group equations.
\end{abstract}

\maketitle
\section{introduction}
Carbon nanotubes are intriguing (quasi-)one-dimensional systems where a variety of electronic states emerge at low temperatures.\cite{saito1998physical, Charlier2007, Ilani2010, Laird2015} Examples are superconductivity in nanotubes embedded in a zeolite matrix,\cite{Tang2001,Lortz2008, Wang2010, Ieong2011, Wang2012} a Mott insulating state in ultraclean nanotubes,\cite{Deshpande2009} and Wigner crystals in semiconducting nanotubes.\cite{Deshpande2008} These diverse physical properties are determined by microscopic details of nanotubes such as wrapping types (armchair, zigzag, or chiral), the number of wrapping (single-wall, multi-wall, or ropes), doping, and correlations. 

In particular, correlations play an important role in presumably metallic nanotubes, since it is well known that one-dimensional gapless systems tend to form (quasi-)long-range order via backscattering processes.\cite{giamarchi2004quantum} The effects of electron-electron (e-e) interactions (both long range and short range) in carbon nanotubes have been extensively studied, \cite{Balents1997,Krotov1997,Kane1997,Egger1997,Lin1998a,Egger1998,Yoshioka1999, Yoshioka1999a, Lin2002, Que2002,Bunder2007, Bunder2008} and various possible phases have been proposed, e.g., Mott insulators, $d$-wave superconductivity, and Luttinger liquids. At the same time, electron-phonon (e-ph) interactions are not negligible, and they may lead to different low temperature phases such as $s$-wave superconductivity,\cite{Sedeki2000,Sedeki2002,Kamide2003,Barnett2005, Carpentier2006, Iyakutti2006} Peierls instability,\cite{Mintmire1992, Huang1996,Dubay2002,Figge2002, Bohnen2004,Connetable2005,Gonzalez2005,Carr2007,Chen2008,Dumont2010,Jakubsky2017} or Wentzel-Bardeen singularity.\cite{DeMartino2004} When both e-e and e-ph interactions coexist, it is not trivial which phase emerges at low temperatures; treating both interactions on an equal footing is of crucial importance.

In this paper, in order to investigate the effect of e-ph interactions against e-e interactions in carbon nanotubes, we employ a renormalization group (RG) method combined with a two-cutoff scaling scheme. This approach enables us to explore low temperature phases without bias. We focus on metallic single-wall nanotubes (SWNTs) with short-range interactions, whose wrapping types are either armchair or zigzag.   Similar calculations have been done, for example, in Refs.~\onlinecite{Sedeki2002, Chen2008}. However, by including all the possible phonon modes and correctly summing up one-loop diagrams for retarded interactions, we reach several new conclusions. First, we find that in armchair nanotubes, as the e-ph coupling becomes strong, eventually Peierls lattice distortion with a wavelength $\sim 1/2 k^0_F$ ($k^0_F$: the Fermi momentum) is induced by phonon softening. Second, this softening is driven by intraband scattering phonon modes, i.e., stretching, radial breathing, or transverse optical modes. We expect that the radial breathing mode or the transverse optical mode is the one that softens in actual nanotubes depending on the radius. Finally, in contrast to armchair nanotubes, zigzag nanotubes do not show a phonon softening instability; the effect of e-ph coupling is insignificant in this case. This is rather unexpected, since the phase diagrams of metallic armchair and zigzag nanotubes are similar in the absence of e-ph coupling. We explain the different behavior of these types of nanotubes based on the RG equations. We show that the structure of RG equations becomes especially simple for carbon nanotubes that we consider, and that the basic analysis of the RG flows can describe the phase diagrams well. 

The paper is organized as follows. In Sec. \ref{Model}, we derive effective low energy models for metallic SWNTs from an extended Hubbard model in graphene. Sec. \ref{RG} presents the phase diagrams obtained by RG analysis with and without electron-phonon interactions. In Sec. \ref{Discussion}, we analyze the RG equations and discuss the consequences of the electron-phonon coupling on low temperature phases. Sec. \ref{Conclusion} is the conclusion. The complete RG equations and other technical details are summarized in the appendices.

\section{Model}
\label{Model}
In this section, we present low energy effective models for SWNTs based on an extended Hubbard model on a graphene lattice. Particular attention is paid to connect the zone-folding scheme, which has been used in Refs.~\onlinecite{Balents1997,Lin1998a, Bunder2007}, and the radial quantization scheme in Ref.~\onlinecite{Mahan2003}. Using the latter is important to correctly derive the electron-phonon coupling in nanotubes, while the former gives a more straightforward interpretation of band structures in nanotubes. The details of the derivation are given in Appendix \ref{AppA}.

Throughout the paper, we assume weak short-range e-e interactions: on-site repulsive interactions $U$, and nearest neighbor interactions $V_{(\perp)}$. In free-standing nanotubes, the long-range Coulomb interaction always exists and it is not small. Weak short-range interactions are realized by screening the Coulomb interaction by, for example, putting SWNTs on a substrate\cite{Hausler2002, Fogler2005} or assembling them in an array.\cite{Gonzalez2005} We, however, expect that the nearest neighbor interactions capture the essential physics induced by the long-range interaction. Indeed, as we demonstrate in the next section, an excitonic insulator phase, which has been discussed in a model with the long-range Coulomb interaction,\cite{Varsano2017} also appears in our model. In experiments, the substrate must be chosen such that the hybridization of nanotube and substrate electrons do not occur; for instance, the surface states of the substrate must be away from the Fermi energy of the nanotubes. Otherwise, the critical temperatures of the ordered phases that we find below will decrease due to the finite life time of quasi-particles.

\subsection{Electronic Hamiltonian}
\begin{figure}[!tb]
\begin{center}
   \includegraphics[width = 0.7\columnwidth ]{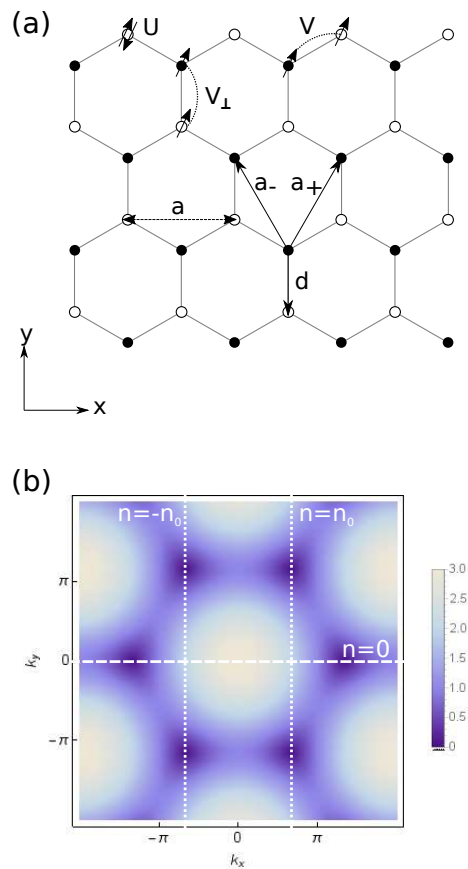}
   \caption{(a) Graphene lattice. Filled (empty) circles represent $A(B)$ sites. (b) The upper band of a graphene tight-binding dispersion in units of $J_0$ in Eq.~\eqref{hk}. The dashed horizontal line corresponds to the gapless mode in armchair nanotubes, and the dotted vertical lines to the ones in zigzag nanotubes.}
\label{Fig1}
\end{center}
\end{figure}

We start from a model of a graphene sheet. A graphene lattice consists of two triangular sublattices, $A$ and $B$ sites, with basis vectors $\bm{a}_{\pm} = a(\pm 1/2, \sqrt{3}/2)$, where $a$ is the distance between neighboring equivalent sites, and the sublattice offset vector $\bm{d} = a(0, - 1/\sqrt{3})$ [Fig. \ref{Fig1}(a)]. The hopping Hamiltonian is 
\begin{equation}\label{h0}
H_0 = -J_0 \sum_{\bm{r} \in \bm{R}, \alpha} \sum_{i=1}^{3} \left[ c^{\dagger}_{A \alpha} (\bm{r})c_{B\alpha} (\bm{r}+\bm{\delta}_i)  + \text{h.c.} \right],
\end{equation}
where $c^{(\dagger)}_{m \alpha}$ is the annihilation (creation) operator of the fermion on sublattice $m$ with spin $\alpha$, and $J_0$ is the hopping energy between neighboring sites. The $A$ sites are at $\bm{R} = n_+ \bm{a}_+  + n_- \bm{a}_-$ with integers $n_{\pm}$, and their neighboring $B$ sites are at $\bm{R} + \bm{\delta}_i$ ($i=1,2,3$) with 
\begin{align}
\bm{\delta}_1 &= \bm{d}, & \bm{\delta}_{2} &= \bm{a}_{-} + \bm{d}, & \bm{\delta}_{3} &= \bm{a}_{+} + \bm{d}. 
\label{bond_dir}
\end{align}
Fourier transforming the hopping Hamiltonian leads to band dispersions $E_{\pm}(\bm{k}) = \pm |h(\bm{k})|$ [see Fig. \ref{Fig1}(b)] with
\begin{equation}
h(\bm{k}) = 2J_0 \cos(k_x a/2) e^{i k_y a /2 \sqrt{3}} + J_0 e^{-i k_y/\sqrt{3}}.
\label{hk}
\end{equation}
An undoped graphene sheet has point-like Fermi surfaces at the Dirac points, where the band dispersion is linear. The on-site interaction is 
\begin{equation}
H_U = U \sum_{\bm{r} \in \bm{R}} \Big[n_{A\up}(\bm{r})n_{A\down}(\bm{r}) + n_{B \up}(\bm{r}+\bm{d})n_{B\down}(\bm{r} +\bm{d}) \Big],
\end{equation}
and the nearest neighbor interactions are
\begin{multline}
H_V = \sum_{\bm{r} \in \bm{R}, \alpha, \beta}\Big[ V_{\perp} n_{A\alpha}(\bm{r})n_{B\beta}(\bm{r}+\bm{\delta}_1) \\
+ V \sum_{i=2,3} n_{A \alpha}(\bm{r})n_{B\beta}(\bm{r} +\bm{\delta}_i)\Big]  .
\end{multline}
We assume that $|V|, |V_{\perp}|  < U$.

We construct a single-wall carbon nanotube by rolling up a graphene sheet. Here we consider $(N_y, N_y)$ armchair carbon nanotubes, and $(N_x, 0)$ zigzag nanotubes. The transverse direction, along which the graphene is wrapped, is the $y$- or $x$-axis for armchair and zigzag nanotubes, respectively. This leads to quantization of momentum in the transverse direction. We move to cylindrical coordinates $\bm{r} = [R_0 \cos(\theta), R_0 \sin(\theta), z]$, where $R_0$ is the radius of the nanotube and $z$ is the position parallel to the tube axis. Now the Fourier transformation is given by 
\begin{equation}
c_{m\alpha} (\bm{r}) = \frac{1}{\sqrt{N_z  N_{\perp}}} \sum_{k,n, \alpha} e^{i (k z+ n \theta)} c_{m k n  \alpha},
\label{Fourier}
\end{equation}
where $N_{z}$ and $N_{\perp}$ are the number of unit cells along the tube and along the radial direction respectively. The quantized angular momentum $n$ is an integer from $-N_{\perp}/2+1$ to $N_{\perp}/2$.\cite{reich2009carbon} The hopping Hamiltonian becomes\cite{Mahan2003}
\begin{equation}
\begin{split}
H_0 &= -J_0 \sum_{k n \alpha} \left[ c_{A k n  \alpha }^{\dagger} c_{B k n  \alpha }\gamma_n (k) + \text{h.c.}\right], \\
\gamma_n (k) &= 
\begin{cases}
e^{i n \theta_1} + 2 \cos(k a /2) e^{-i n \theta_1 /2} & \text{armchair} \\
e^{-i k a /\sqrt{3}} + 2 e^{i k a /2\sqrt{3}} \cos(n \theta_z/2) & \text{zigzag}
\end{cases},
\end{split}
\label{gamma}
\end{equation}
where $\theta_1$ and $\theta_z$ are the angles between sites along the circumference in armchair and zigzag geometries;\cite{Mahan2003}  $\theta_1 \simeq a/\sqrt{3}R_0$ and $\theta_z  \simeq a/R_0$. The eigenvalues are $E_{\pm} (k,n) = \pm J_0 |\gamma_n (k)|$. There always exists a gapless mode for armchair nanotubes, i.e., $n=0$. On the other hand, the zigzag nanotubes have gapless modes only when $N_x$ is an integer multiple of three, and their gapless modes are with $n = n_0 \equiv N_x/3$. These gapless modes correspond to lines in the original graphene dispersion, and they are depicted in Fig.~\ref{Fig1}(b).

\begin{figure}[!tb]
\begin{center}
   \includegraphics[width = 1.0\columnwidth ]{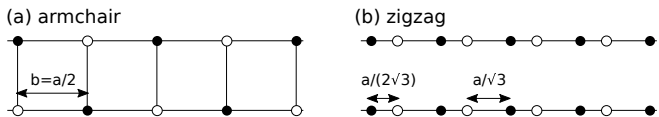}
   \caption{Effective two-leg ladder models obtained by considering only the gapless modes of the tight-binding Hamiltonian for armchair and zigzag nanotubes.}
\label{two-leg}
\end{center}
\end{figure}

We take into account only the gapless modes since we are interested in low energy physics. For armchair and zigzag nanotubes, the effective low energy models reduce to two-leg ladder models (Fig.~\ref{two-leg}).\cite{Balents1997,Lin1998a, Bunder2007} The detail of the derivation is given in Appendix ~\ref{AppA}. For armchair nanotubes, the $n=0$ mode of the diagonalized hopping Hamiltonian gives
\begin{equation}
H_0 = \sum_{q = \pm} \sum_{k, \alpha} E_{q} (k) d^{\dagger}_{q k \alpha} d_{q k \alpha},
\end{equation}
with $E_{\pm} (k) = - J_0 \cos(k b) \mp J_0 $. The effective lattice constant is $b = a/2$ [Fig.~\ref{two-leg}(a)], and the new Brillouin zone is $[-\pi/b, \pi/b]$. We label the two bands with $q=\pm$; each has two Fermi points of right and left moving branches. For zigzag nanotubes, the $n=\pm n_0$ modes of the diagonalized hopping Hamiltonian give 
\begin{equation}
H_0 = \sum_{q = \pm} \sum_{r = R, L} \sum_{\alpha} E_{r} (k) d^{\dagger}_{r q k \alpha} d_{r q k \alpha},
\end{equation}
with $E_{R/L} = \mp 2 J_0 \cos(kb')$. $b' = a \sqrt{3}/4$ is the average distance between neighboring sites, and the unit cell size is now $2b'$ [Fig.~\ref{two-leg}(b)]. The new Brillouin zone is taken as $[0, \pi/b']$. We have four bands labeled by $(q, r)$. Here $q=\pm$ correspond to $n=\pm n_0$, and $r=R,L$ denotes the chirality of the band near the Fermi energy. Each band has one Fermi point near $\pi/b'$. When the system is undoped, the two bands in the armchair nanotubes have the same Fermi velocity, while they are no longer the same for doped cases. Instead, the zigzag nanotubes have always doubly degenerate bands ($q=\pm1$), and their Fermi velocities are the same. In the following, we always consider either no doping or infinitesimal doping regimes. Therefore, we ignore the velocity difference between two bands, and the additional gapless modes that may appear for doped systems. 

We note that the curvature effect on the electronic dispersions becomes important for small radius nanotubes;\cite{Blase1994, Balents1997, Milosevic2004, reich2009carbon}  the Dirac points are slightly shifted by an amount of order $R_0^{-2}$. This is not a problem for armchair nanotubes, since the Dirac points still coincide with the quantized radial momenta keeping the system metallic. On the other hand, the Dirac points in zigzag nanotubes shift away from the quantized momenta inducing a small energy gap $\propto R_0^{-2}$; the system becomes semi-conducting. However, it is possible that such a small energy gap may not affect the following analysis in the presence of interactions, which scales as $R_0^{-1}$.\cite{Balents1997} Therefore we will ignore the curvature effect in the following.

To construct low energy effective models, we take the continuum limit,
\begin{align}
b^{(\prime)} &\sum_z \rightarrow \int dz, & \frac{1}{\sqrt{b^{(\prime)}}} d (z) \rightarrow \psi (z).
\end{align}
We then linearize the dispersion around the Fermi energy. This introduces two chiral fields $\psi_{R, L} (z)$, which vary slowly compared to $1/k_F$. The kinetic term is
\begin{equation}
H_0 = v \sum_{q, \alpha} \int dz  \left(\psi_{R q \alpha}^{\dagger} i \partial_z \psi_{R q \alpha} - \psi_{L q \alpha}^{\dagger} i \partial_z \psi_{L q \alpha} \right),
\end{equation}
where $v$ is the Fermi velocity. After substituting the chiral decomposition of the low energy modes [Eqs.~\eqref{chiral1} and \eqref{chiral2}] into the interaction part of the Hamiltonian, we can cast both cases into the following form,  
\begin{equation}
\begin{split}
\frac{H_\text{int}}{2 \pi v} &= c^s_{qq'}\psi^{\dag}_{Rq\alpha}\psi^{\dag}_{Lq\beta}\psi_{Lq'\beta}\psi_{Rq'\alpha}\\
&+c^l_{qq'}\psi^{\dag}_{Rq\alpha}\psi^{\dag}_{Lq\beta}\psi_{Rq'\beta}\psi_{Lq'\alpha}\\
&+f^s_{qq'}\psi^{\dag}_{Rq\alpha}\psi^{\dag}_{Lq'\beta}\psi_{Lq'\beta}\psi_{Rq\alpha}\\
&+f^l_{qq'}\psi^{\dag}_{Rq\alpha}\psi^{\dag}_{Lq'\beta}\psi_{Rq\beta}\psi_{Lq'\alpha}\\
&+\frac{1}{2}u^+_{qq'}\left(\psi^{\dag}_{Rq\alpha}\psi^{\dag}_{Rq'\beta}\psi_{L\bar{q}\beta}\psi_{L\bar{q'}\alpha}+\rm{h.c.}\right )\\
&+\frac{1}{2}u^-_{qq'}\left( \psi^{\dag}_{Rq\alpha}\psi^{\dag}_{Rq'\beta}\psi_{L\bar{q}\alpha}\psi_{L\bar{q'}\beta}+\rm{h.c.}\right),
\end{split}
\label{Hint}
\end{equation}
where Einstein summation is implicitly assumed over $q^{(\prime)} = \pm = 1,2$. In order to avoid double counting, we choose $f_{qq}=0$ and $u^-_{qq}=0$. Furthermore, due to hermiticity, we have $c_{q'q}=c_{qq'}$ and $u_{qq'}=u_{\bar{q}\bar{q'}}$, and the parity symmetry implies $f_{q'q}=f_{qq'}$. Thus, there are nine independent coupling constants: $( c_{11}^l, c_{11}^s,c_{12}^l, c_{12}^s,f_{12}^l, f_{12}^s, u_{11}^+, u_{12}^+, u_{12}^-  )$. The initial (bare) values of these coupling constants in terms of $U$, $V$ and $V_{\perp}$ are given in Appendix \ref{AppB}. Umklapp processes, which are described by the parameters $u_{qq'}^{+/-}$, are absent for doped nanotubes. 

Finally, using the operator product expansion,\cite{Balents1996, cardy1996scaling, Delft1998} we calculate the RG equations for these coupling constants. To identify the emerging order, we consider the renormalized bosonized theory, as in Refs.~\onlinecite{giamarchi2004quantum, gogolin2004bosonization, Delft1998, Haldane1981, Voit1995, Carpentier2006}. In many cases, RG analysis combined with bosonization gives reliable results in one-dimension even when coupling constants flow to the strong coupling limit, i.e., asymptotic free theory.\cite{Chen2004,Chang2005} Here we only cite the final form,
\begin{multline}
\frac{H_\text{int}}{2 \pi v}= \frac{1}{(\pi \alpha_0)^2}\Big[ c_{11}^l \cos (2 \phi_{s0}) \cos (2 \phi_{s\pi}) \\
\begin{split}
&- c_{12}^l \cos (2 \theta_{c\pi}) \cos (2 \phi_{s0}) -  c_{12}^s \cos (2 \theta_{c\pi}) \cos (2 \phi_{s\pi}) \\
&- \left( c_{12}^l - c_{12}^s\right) \cos (2 \theta_{c\pi}) \cos (2 \theta_{s\pi}) \\
&- f_{12}^l \cos (2 \phi_{s0}) \cos (2 \theta_{s\pi}) - u_{11}^+ \cos (2 \phi_{c0}) \cos (2 \theta_{c\pi}) \\
&- u_{12}^+ \cos (2 \phi_{c0}) \cos (2 \phi_{s\pi}) - u_{12}^- \cos (2 \phi_{c0}) \cos (2 \theta_{s\pi}) \\
&- \left( u_{12}^+ +u_{12}^- \right) \cos (2 \phi_{c0}) \cos (2 \theta_{s 0 })\Big],
\end{split}
\label{bosonization}
\end{multline}
and the detailed derivation is given in Appendix \ref{AppC}. When the system is undoped, i.e., at commensurate filling, all four modes are gapped and pinned. The possible phases and their pinned fields are given in Table \ref{Table1}, and also discussed in Refs.~\onlinecite{Balents1996,Lin1998,Azaria2000, Tsuchiizu2002a,Wu2003, Chudzinski2008, Nonne2010, Okamoto2012}. The properties of these phases are explained in the next section. For incommensurate filling, the total charge mode $\phi_{c0}$ becomes massless. Then, Mott phases become superconducting phases with the same local pairing symmetry. At the same time, the charge-density wave (CDW) phase and the $p$-wave density wave (PDW) phase become degenerate since the $Z_2$ symmetry ($\braket{\phi_{c0}} = 0$ or $\pi$) is unbroken. Similarly, the chiral current phase (CCP) and $f$-wave density wave (FDW) become degenerate.   
\begin{table}[!tb]
\caption{Expectation values of bosonic variables in the fully gapped phases. We set $\langle \phi_{c0}\rangle =0$.}
\begin{ruledtabular}
\centering
\begin{tabular}{cccccc}
Phase &  $\langle \phi_{c\pi}\rangle$ &  $\langle \phi_{s0}\rangle$ & $\langle \phi_{s\pi}\rangle$ & $\langle \theta_{s\pi}\rangle$ \\
\hline
CDW &  $\pi/2$ & 0 &&0\\
PDW &   0 & $\pi/2$ &&$\pi/2$\\
CCP & 0 &0&&0  \\
FDW &$\pi/2$ &$\pi/2$ &&$\pi/2$\\
$S$-Mott  &$\pi/2$ & 0 &0&\\
$S'$-Mott   & 0 & $\pi/2$ &$\pi/2$&\\
$D$-Mott &0 &0&0& \\
$D'$-Mott&$\pi/2$ &$\pi/2$ &$\pi/2$&
\end{tabular}
\label{pinned}
\end{ruledtabular}
\label{Table1}
\end{table}

\subsection{Electron-phonon interactions}
Since a full microscopic description of the e-ph interaction and its parameters are not accessible, we follow the treatment by Mahan in Ref.~\onlinecite{Mahan2003}, and introduce the electron-phonon coupling as hopping modulations,
\begin{multline}
V_\text{e-ph} = -J_1 \sum_{\bm{r} \in \bm{R}, \alpha} \sum_{i=1}^3 \hat{\bm{\delta}}_i \cdot \left[\bm{Q}_{B,i} (\bm{r}) - \bm{Q}_A (\bm{r})\right] \\
\times \left[ c^{\dagger}_{A \alpha} (\bm{r})c_{B\alpha} (\bm{r}+\bm{\delta}_i)  + \text{h.c.} \right],
\end{multline}
where $\bm{Q}_A (\bm{r})$ and $\bm{Q}_{B, i} (\bm{r})$ are the lattice displacement vectors for an $A$ site and its surrounding $B$ sites. $\hat{\bm{\delta}}_i$ is the normalized bond vectors given in Eq.~\eqref{bond_dir}. Fourier transforming $\bm{Q}_{A}$ and $\bm{Q}_{B,i}$ as Eq.~\eqref{Fourier}, we obtain
\begin{equation}
V_\text{e-ph} = -J_1 \sum_{k k' n n' \alpha} \left[ M_{k,k'}^{n,n'} c^{\dagger}_{A, k+k', n+n', \alpha} c_{B k n \alpha}  + \text{h.c.}\right],
\end{equation}
where $M_{k,k'}^{n,n'}$'s are linear combinations of $\bm{Q}_{m,k'}$. The exact expressions can be found in Ref.~\onlinecite{Mahan2003}. Each displacement vector consists of radial, transverse, and longitudinal (along the tube axis) components, $(Q_{m\rho}, Q_{m\theta}, Q_{m z})$. In total there are six modes, and it is convenient to use a new basis: $Q_{\nu} = (Q_{A\nu} + Q_{B\nu})/2$ and $q_{\nu} = Q_{A\nu} - Q_{B\nu}$. The first three are denoted as acoustic modes, and the other three are denoted as optical modes.\footnote{$Q_{\rho}$ is called the radial acoustic mode since its frequency is zero at zero momentum in graphene. In carbon nanotubes. the frequency becomes nonzero. See, for example, Ref.~\onlinecite{saito1998physical}.} Three acoustic modes ($Q_{\rho}$, $Q_{\theta}$, $Q_z$) are often called as breathing, twisting, and stretching modes respectively.

For armchair nanotubes, we only consider $n = n' = 0$. Moving to the eigenstate basis, we find \cite{Mahan2003} 
\begin{equation}
\begin{split}
V_\text{e-ph} &= V_1 + V_2,\\
V_1 &= \sum_{k,k', \alpha} \sum_{\nu = \text{LA,RA,TO}}g_{\nu p} (G)\left(b_{p \nu} + b^{\dagger}_{-p, \nu}\right) \\
&\times \left(d^{\dagger}_{+k' \alpha} d_{+ k \alpha}- d^{\dagger}_{-, k'+\pi, \alpha} d_{-,k +\pi, \alpha}\right), \\
V_2 &= \sum_{\nu, k,k', \alpha} \sum_{\nu' = \text{TA,LO,RO}} g_{\nu p} (G)\left(b_{p \nu} + b^{\dagger}_{-p, \nu}\right) \\
&\times  \left(d^{\dagger}_{+k' \alpha} d_{-, k+\pi, \alpha}- d^{\dagger}_{-,k'+\pi, \alpha} d_{+ k \alpha}\right),
\end{split}
\label{eph1}
\end{equation}
where $V_1$ is the intraband scattering caused by longtitudinal acoustic (LA), radial acoustic (RA), and transverse optical (TO) modes, and $V_2$ is the interband scattering caused by transverse acoustic (TA), longitudinal optical (LO), and radial optical (RO) modes. The reciprocal vector $G$ is taken such that $p=k-k'+G$ lies within the first Brillouin zone.\cite{Sedeki2000, bruus2004many} The displacement vectors $Q_{\nu}$ are quantized and $g_{\nu}$ is the coupling constants for mode $\nu$. 

For zigzag nanotubes, we approximate $M_{k,k'}^{n,n'}  \simeq M_{\pi/2b', 0}^{n,n'}$ with $n, n+n'= \pm n_0$, since we only consider states near the Fermi points located at or very close to the Brillouin zone boundary. In the eigenstate basis, we find
\begin{equation}
\begin{split}
V_\text{e-ph} &= V_1 + V_2,\\
V_1 &\simeq \sum_{k, \alpha, q}' \sum_{\nu' = \text{RA,LO}} g_{\nu,0} \left(b_{0, \nu} + b^{\dagger}_{0, \nu}\right) \\
&\times  \left(d^{\dagger}_{L q k \alpha} d_{R q k \alpha} - d^{\dagger}_{R q k \alpha} d_{L q k  \alpha}\right),\\
V_2 &\simeq \sum_{k, \alpha, q}' \sum_{\nu' = \text{RA, TA, LO}} g_{\nu, 2n_0} \left(b_{2n_0, \nu} + b^{\dagger}_{-2n_0, \nu}\right) \\
&\times  \left(d^{\dagger}_{L \bar{q} k \alpha} d_{R q k  \alpha}- d^{\dagger}_{R \bar{q} k \alpha} d_{L q k \alpha}\right).
\end{split}
\label{eph2}
\end{equation}
The summation over $k$ is restricted near the Fermi surface. $V_1$ is the intraband scattering, where the momentum or angular momentum transfer by phonons is negligible. $V_2$ is the interband scattering, where an electron is scattered from one branch to the other one. We ignore an additional scattering process by TO phonons
\begin{multline}
V_3 \simeq \sum_{k, \alpha, q}' q g_\text{TO}\left(b_{0, \text{TO}} + b^{\dagger}_{0, \text{TO}}\right)\\
 \times \left(d^{\dagger}_{R q k \alpha} d_{R q k \alpha} - d^{\dagger}_{L q k \alpha} d_{Lq k \alpha} \right),
\end{multline}
since this effectively renormalizes the $f_{12}^s$ term, which is irrelevant in the renormalization group equations below.

In the next section, we will derive the effective retarded interaction among electrons by integrating out the phonons. This treatment and the use of a two-cutoff scaling scheme are phenomenological by nature.\cite{Caron1984} Therefore, we do not elaborate on the precise values of e-ph coupling constants here.

\section{Renormalization group analysis}
\label{RG}

\subsection{Without e-ph interactions}
RG equations for general $N$-leg ladder problems with instantaneous electronic interactions have been discussed in Refs.~\onlinecite{Lin1997,Chang2005}. For the sake of completeness, these equations are cited in Appendix~\ref{AppB}. 

\begin{figure}[!tb]
\begin{center}
   \includegraphics[width = 1.0\columnwidth ]{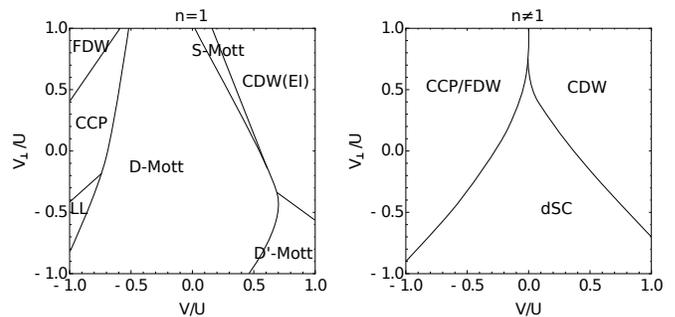}
   \caption{Weak-coupling phase diagrams of armchair nanotubes without e-ph interactions for undoped (left) and doped (right) systems. }
\label{PD_arm}
\end{center}
\end{figure}

\begin{figure}[!tb]
\begin{center}
   \includegraphics[width = 1.0\columnwidth ]{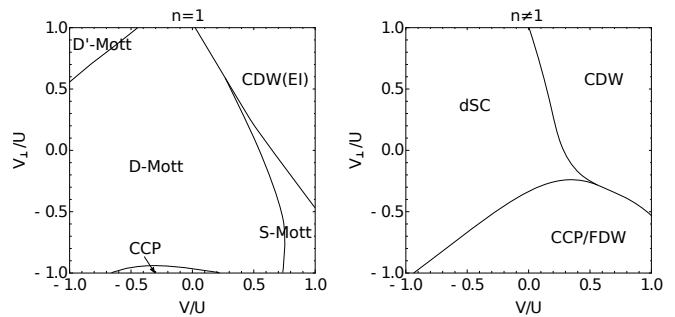}
   \caption{Weak-coupling phase diagrams of zigzag nanotubes without e-ph interactions for undoped (left) and doped (right) systems. }
\label{PD_zigzag}
\end{center}
\end{figure}

\begin{figure}[!tb]
	\begin{center}
		\includegraphics[width = \columnwidth ]{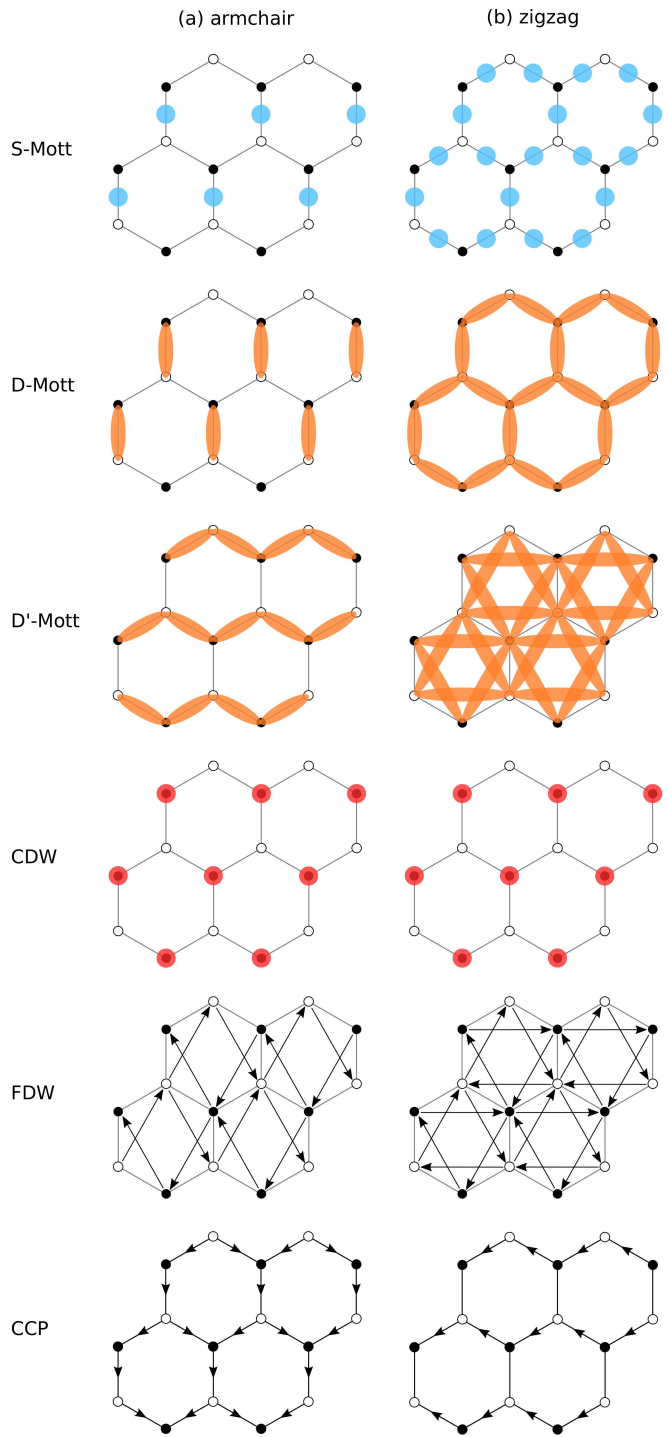}
		\caption{Schematic presentation of phases appearing in armchair and zigzag nanotubes at half-filling. Blue circles represent the $s$-wave pairing, orange ellipses the $d$-wave pairing, red dots charge accumulations, and arrows currents (We follow the notation of Ref.~\onlinecite{Bunder2008}).}
		\label{OP}
	\end{center}
\end{figure}

The phase diagrams for vanishing e-ph interactions for armchair and zigzag nanotubes are shown in Figs.~\ref{PD_arm} and \ref{PD_zigzag}. Corresponding order parameters are sketched in Fig.~\ref{OP} following the notation of Ref.~\onlinecite{Bunder2008}. The phase diagrams are obtained by integrating the RG equations with initial conditions $U/J_0 = 10^{-8}-10^{-5}$ until one or several of the coupling constants reach the order of unity. We use such small initial conditions to focus on the asymptotic diverging flows that can be well captured by the ansatz \cite{Balents1996, Chen2004, Seidel2005, Chang2005, Cai2012}
\begin{equation}
g_{i}(l) \sim \frac{G_i}{l_d-l},
\label{ansatz}
\end{equation}
where $l$ gives the running ultraviolet cutoff $\tilde{\Lambda} = \Lambda_0 e^{-l}$. We analyze the effective model in bosonized form to determine the low energy phases. The pinned values of the bosonic fields and the corresponding phases are given in Table~\ref{Table1}. The phase diagrams for zigzag nanotubes have been obtained in Ref.~\onlinecite{Bunder2008}. We note that the two phase diagrams are mapped to each other via the following transformation,
\begin{equation}
V  \leftrightarrow \frac{1}{3} (V + 2 V_{\perp}), \  V_{\perp} \leftrightarrow \frac{1}{3} (4V - V_{\perp}).
\end{equation}
This approximately exchanges $V$ and $V_{\perp}$, or equivalently the $x$- and $y$- axes in Fig.~\ref{Fig1}(a). Since zigzag/armchair nanotubes are rolled up along the $x/y$-axis, such exchange of $V$ and $V_{\perp}$ interchanges the phase diagrams within the bosonization language. However, real space order parameters are different for two nanotubes even for the same pinning fields as we discuss below (see Fig.~\ref{OP}).

First, in both types of SWNTs, we find various Mott phases for the undoped case.\cite{Tsuchiizu2002a, Bunder2008} These Mott phases have on average one electron per site, and no hopping is allowed due to repulsive interactions. They become superconducting states when the system deviates from half-filling, and have local pairing that resemble their partner superconducting states. In bosonization language, the only difference between the Mott states and their superconducting partners is whether the total charge mode, $\phi_{c0}$, is gapped or not. The local pairing in the $s$-wave Mott phases consists of on-site pairs
\begin{equation}
\frac{1}{\sqrt{2}} \left(  \ket{\up \down, 0} + \ket{0,\up \down} \right),
\end{equation}
while the pairing in the $d$-wave Mott phases consists of spin singlets 
\begin{equation}
\frac{1}{\sqrt{2}} \left(  \ket{\up, \down} - \ket{\down, \up} \right).
\end{equation}
In armchair nanotubes, the transverse and tube directions are not equivalent, and the local pairs are on the transverse bonds in the $S$- and $D$-Mott phases, while the $S'$- and $D'$-Mott states have pairing along the tube direction (Fig.~\ref{OP}). On the other hand, in zigzag nanotubes, local pairs are distributed equally on every nearest neighbor bonds in the $S$- and $D$-Mott phases, and on every next-nearest neighbor bonds in the $S'$- and $D'$-Mott phase  (Fig.~\ref{OP}). 

The CDW phase appears when $V, V_{\perp} \sim U$ as observed in Ref.~\onlinecite{Tsuchiizu2002a}. For the undoped case, this corresponds to a charge modulation on $A$ and $B$ sites, 
\begin{align}
n_A & = 1 \pm \delta n , & n_B &= 1 \mp \delta n.  
\end{align}
In this regard, this CDW state is similar to the excitonic insulator (EI) phase, which was proposed for metallic carbon nanotubes with long-range interactions.\cite{Varsano2017} We expect that long-range interactions beyond nearest neighbor repulsions further stabilize the CDW phase. When the system is doped, the CDW is no longer commensurate, and the periodicity of the charge modulation along the tube axis becomes $1/(k_F^1+k_F^2)$ for armchair nanotubes and $1/(2 k_F)$ for zigzag nanotubes.

Chiral current phases (CCPs),\cite{Bunder2008} where time-reversal symmetry is spontaneously broken, appear in both phase diagrams. Similar chiral states have been discussed in cold atom gases,\cite{Olschlager2011,Wirth2011,Soltan-Panahi2012,Olschlager2013} superfluid $^3$He,\cite{Leggett1975} or Sr$_2$RuO$_4$.\cite{Mackenzie2003} The CCP is mapped on the staggered flux (SF) phase in two-leg ladder problems, which was proposed in the context of $t$-$J$ model.\cite{Affleck1988,Marston1989, Marston2002, Tsuchiizu2002a} The SF phase has circularly flowing currents on a plaquet of a ladder, and its chirality is spatially staggered along the ladder. However, we note that the CCPs in the two types of SWNTs look different in the original honeycomb lattice, since there are no ``staggered currents" there. In SWNTs, the current circulates along the circumference of the tubes in either a clockwise or anti-clockwise manner; they look like a solenoid (Fig.~\ref{OP}).
    
The new phases that appear in the armchair case, albeit in unphysical parameter regimes, are an $f$-wave density wave (FDW) and a Luttinger liquid (LL). The FDW state has a staggered current,
\begin{multline}
J = i \sum_{\alpha} \Big[ d^{\dagger}_{A\alpha} (z+b) d_{A\alpha} (z)- d^{\dagger}_{B\alpha} (z+b) d_{B\alpha} (z) - \text{h.c.} \Big].
\end{multline}
The FDW and the CCP become degenerate as the system is doped. In the Luttinger liquid phase, all four bosonic modes are massless until $l \sim 10^{10}$, where we stop our RG flow.

\subsection{With e-ph interactions}
In the presence of phonons, there is another energy scale in addition to the ultraviolet cutoff, i.e., the Debye frequency $\omega_\text{D}$. Here we employ a two-cutoff scaling scheme to treat the problem. \cite{Caron1984,Kivelson1988,Zimanyi1988, Sedeki2002, Seidel2005,Cai2012} We first integrate out the phonon modes in Eqs.~\eqref{eph1} and \eqref{eph2} to obtain an effective retarded interaction among electrons \cite{bruus2004many}
\begin{equation}
V_\text{eff} = - 2 |g_{\nu p}|^2 \frac{\omega_p^{\nu}}{\omega_p^{\nu ,2} - \omega^2}.
\label{Veff}
\end{equation}
For $\omega \gg \omega_{p}^{\nu}$, we can ignore the effective interactions. For $\omega \rightarrow 0$, the above expression becomes nearly constant for both acoustic and optical phonons as discussed below. Thus we can approximate it as
\begin{equation}
V_\text{eff} \simeq -2 |g_{\nu p}|^2 \omega^{-1}_\text{D} \Theta(\omega_\text{D} - \omega).
\label{Veff2}
\end{equation}
This resembles the treatment of the phonon-induced interaction in BCS theory. For optical phonons, which are approximately dispersionless, the effective interaction in Eq.~\eqref{Veff} is nearly constant, $-2 |g_{\nu p}|^2/\omega_\text{D}$, below the phonon frequency $\omega_\text{D}$. For acoustic phonons, the coupling constant $|g_{\nu p}|^2$ vanishes linearly as $p \rightarrow 0$, and thus we can approximate the phonon frequency by the zone-boundary values.\footnote{Since each term in Eqs. \eqref{eph1} and \eqref{eph2} includes optical phonons, there is always a finite scattering probability even for $p=0$.} By identifying this frequency as the Debye frequency, we recover the expression in Eq.~\eqref{Veff2}. For the sake of simplicity, we use a single Debye frequency in the following. This gives frequency dependent coupling constants $g_i (\omega)$
\begin{equation}
g_i(\omega) = g_i + \Theta (\omega_D - \omega) \tilde{g}_i,
\end{equation}
where $g_i$ is the instantaneous electronic coupling and $\tilde{g}_i$ is the phonon mediated retarded interactions from $V_\text{eff}$ above. 

In the two-cutoff scaling scheme, RG equations for retarded interactions are given by loop diagrams, since phonon-mediated interactions $\tilde{g}_i$ can only transfer energies $\omega < \omega_\text{D}$. These one loop diagrams must include at least one vertex of $\tilde{g}_i$ due to the step-like energy cutoff $\Theta$. The instantaneous interactions $g_i$ are renormalized only through instantaneous vertices and thus are independent of $\tilde{g}_i$. The one-loop diagrams that we consider include not only the RPA bubbles,\cite{Caron1984, Sedeki2002} but also half-bubble diagrams.\cite{Kivelson1988,Zimanyi1988, Seidel2005, Cai2012} The retarded coupling constants that enter the RG equations are: $(\tilde{c}_{11}^l, \tilde{c}_{12}^l, \tilde{f}_{12}^l, \tilde{u}_{11}^+, \tilde{u}_{12}^+, \tilde{u}_{12}^-)$. The RG equations of these are given in Appendix~\ref{AppB}.

Since the RG equations for the instantaneous interaction parameters $g_i$ are independent of the retarded interaction parameters $\tilde{g}_i$, the former still diverge at length scale $l_d$ as Eq.~\eqref{ansatz}. On the other hand, the retarded interactions $\tilde{g}_i$ may have another length scale $l_p$ at which their RG flow diverge. Since $g_i$ are included in the RG equations for $\tilde{g}_i$, the two length scales should follow $l_p \leq l_d$.  Now what kind of fixed point appears depends on the two length scales. Below we investigate the effect of e-ph coupling to the physically relevant phases for $V = V_{\perp} > 0$.

\subsubsection{Armchair nanotubes}
\begin{figure}[!tb]
\begin{center}
   \includegraphics[width = 1.0\columnwidth ]{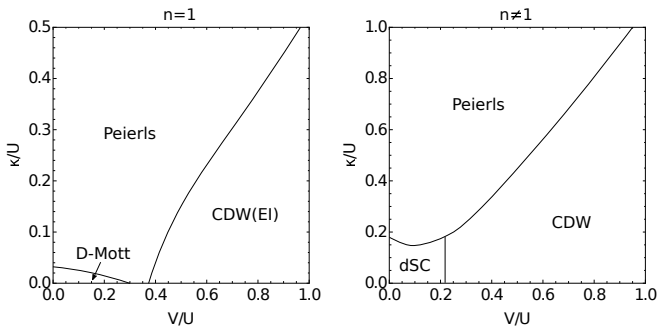}
   \caption{Phase diagrams for armchair nanotubes with e-ph coupling for undoped (left) and doped (right) cases. We take $\kappa_1 = \kappa_2 = \kappa$ in Eq.~\eqref{kappa1}.}
\label{PD2}
\end{center}
\end{figure}

The initial conditions for the retarded coupling constants are $(\tilde{c}_{11}^l, \tilde{c}_{12}^l, \tilde{f}_{12}^l, \tilde{u}_{11}^+, \tilde{u}_{12}^+, \tilde{u}_{12}^-) = 2(-\kappa_1, -\kappa_2, \kappa_2, -\kappa_2, \kappa_1, -\kappa_2)$, where 
\begin{equation}
\begin{split}
\kappa_1 &= 2\omega_\text{D}^{-1} \sum_{\nu = \text{LA,RA,TO}} |g_{\nu, 2k_F}|^2 , \\
\kappa_2 &= 2\omega_\text{D}^{-1} \sum_{\nu = \text{TA,LO,RO}} |g_{\nu, 0}|^2 .
\end{split}
\label{kappa1}
\end{equation}
We consider $\kappa_1 = \kappa_2 = \kappa$ and the phase diagram in terms of $\kappa$ and $V = V_{\perp}$ is given in Fig.~\ref{PD2}. The $S$-Mott phase between $D$-Mott and CDW phases immediately disappears as $\kappa$ is turned on. We find that when $\kappa$ becomes strong, eventually $-\tilde{c}_{11}^l$ and $\tilde{u}_{12}^+$ flow to $+\infty$, or in other words $\kappa_1 \rightarrow + \infty$. From Eq.~\eqref{kappa1}, this implies that the intraband phonons are softened.\cite{Caron1984} The phonon softening leads to a Peierls lattice distortion with periodicity $\sim 1/2k^0_F \sim  3a/8\pi$. 

Because $\kappa$ contains contributions of several phonon modes, the divergence of $\kappa$ does not indicate which of these modes softens. Considering the coupling strength and the mode frequencies,\cite{Barnett2005} we speculate that the RA mode, i.e., radial breathing mode, is the one that softens for a relatively small radius $\sim 4$\AA{}. Considering that the coupling strength of the RA mode and its frequency scales as $R_0^{-2}$ and $R_0^{-1}$ respectively,\cite{Suzuura2002, Mahan2003, Machon2005, Nikolic2010} we expect that the TO mode, i.e., in-plane optical mode, becomes more dominant for larger radius nanotubes. For doped nanotubes, the $d$-wave superconducting phase and the CDW phase turn into Peierls states as $\kappa$ becomes large, while they are more stable than the corresponding phases at half-filling. As we will see in the next section, $\kappa_1$, which triggers the Peierls instability, is always more relevant than $\kappa_2$. Thus, the phase diagram does not change significantly when we deviate from $\kappa_1 = \kappa_2$.

\subsubsection{Zigzag nanotubes}
The initial conditions for the retarded coupling constants are $(\tilde{c}_{11}^l, \tilde{c}_{12}^l, \tilde{f}_{12}^l, \tilde{u}_{11}^+, \tilde{u}_{12}^+, \tilde{u}_{12}^-) = 2(\kappa_4, \kappa_3, \kappa_3, -\kappa_3, -\kappa_4, \kappa_3)$, where 
\begin{equation}
\begin{split}
\kappa_3 &= 2\omega_\text{D}^{-1} \sum_{\nu = \text{LA,RA,TO}} |g_{\nu, 2k_F}|^2 , \\
\kappa_4 &= 2\omega_\text{D}^{-1} \sum_{\nu = \text{TA,LO,RO}} |g_{\nu, 0}|^2 .
\end{split}
\end{equation}
In this case, we do not find additional instabilities induced by phonons for $\kappa_{3,4} < U$. This indicates that the phases induced by electronic interactions in zigzag nanotubes are fairly stable to phononic perturbations. We discuss why zigzag nanotubes are much more stable than armchair nanotubes by analyzing the RG equations in detail in the next section.

\section{Discussion}
\label{Discussion}
In this section we analyze the RG equations for the retarded interactions to clarify the different behavior of armchair and zigzag nanotubes. First, we decouple these RG equations into the following form, 
\begin{equation}
\dot{h}_i = h_i \rho_i - 2h_i^2  ,
\label{Seidel}
\end{equation}
by introducing new variables $h_1, \dots, h_6$,
\begin{equation}
\begin{split}
\begin{pmatrix}
 h_1\\ 
 h_2 
\end{pmatrix} 
&=
\begin{pmatrix}
1 & 1   \\ 
 1& -1 
\end{pmatrix} 
\begin{pmatrix}
 \tilde{c}^l_{11}\\ 
 \tilde{u}^+_{12}
\end{pmatrix}, \\
\begin{pmatrix}
 h_3\\ 
 h_4 \\ 
 h_5 \\ 
h_6
\end{pmatrix} 
&= 
\begin{pmatrix}
-1 & 1 & -1  &-1  \\ 
 -1& 1 &1  &1  \\ 
 1& 1 &1  &-1  \\ 
 1& 1 & -1  &1 
\end{pmatrix} 
\begin{pmatrix}
 \tilde{c}^l_{12}\\ 
 \tilde{f}^l_{12} \\ 
 \tilde{u}^+_{11} \\ 
 \tilde{u}^-_{12} 
\end{pmatrix}.
\end{split}
\end{equation}
The parameters $\rho_i$ are linear combinations of instantaneous coupling constants,
\begin{equation}
\begin{split}
\rho_{1,2} &= - 2\left( 2 c_{11}^l - c_{11}^s \pm 2u_{12}^+ \pm u_{12}^- \right), \\
\begin{pmatrix}
 \rho_3\\ 
 \rho_4 \\ 
 \rho_5 \\ 
\rho_6
\end{pmatrix} 
&= 2
\begin{pmatrix}
1 & 1 & 1  &1  \\ 
 1& 1 &-1  &-1  \\ 
 -1& 1 &-1  &1  \\ 
 -1& 1 & 1  &-1 
\end{pmatrix} 
\begin{pmatrix}
 2 c_{12}^l - c_{12}^s\\ 
 -2 f_{12}^l + f_{12}^s \\ 
 u_{11}^+ \\ 
 u_{12}^+ + 2u_{12}^- 
\end{pmatrix} .
\end{split}
\end{equation}

\begin{figure}[!tb]
	\begin{center}
		\includegraphics[width = 0.6\columnwidth ]{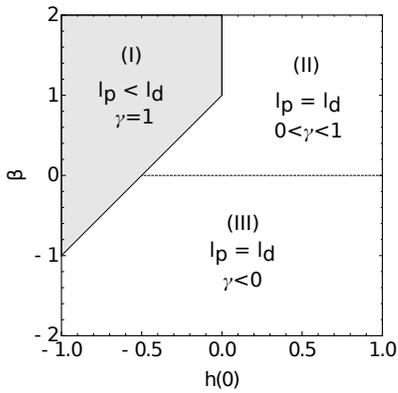}
		\caption{Parameter dependence of the asymptotic solutions in Eq.~\eqref{tilde_asym}. Among the three possible behaviors, only the region I gives the phonon-driven distinct phases.}
		\label{Seidel_fig}
	\end{center}
\end{figure}

Eq.~\eqref{Seidel} has the same form as the RG equation analyzed in Ref.~\onlinecite{Seidel2005}. Its formal solution is given as
\begin{equation}
h_i (l) = M_i(l) \left[ \int_0^l dl' 2 M_i(l') + h_i (0)^{-1}\right]^{-1},
\end{equation}
with $M_i(l) \equiv \exp[ \int_0^l dl' \rho_i(l') ]$. Since the instantaneous coupling diverges as Eq.~\eqref{ansatz}, $\rho_i$ behaves as
\begin{equation}
\rho_i \simeq \frac{\beta_i}{ l_d - l}.
\label{rhos}
\end{equation}
The RG flow of $h_i$ is roughly determined by the value of $\beta_i$ and the initial value $h_i (0)$.\cite{Seidel2005} The asymptotic behavior of $h_i(l)$ are well captured by an ansatz
\begin{equation}
h_i(l) \sim \frac{\tilde{G}_i}{(l_p - l)^{\gamma_i}}.
\label{tilde_asym}
\end{equation}
In Fig.~\ref{Seidel_fig}, we sketch the three regimes given by $\beta_i$ and $h_i (0)$. In region I, the retarded coupling diverges faster than the instantaneous coupling, i.e., $l_p < l_d$ with $\gamma = 1$. On the other hand, in region II, the retarded couplings are diverging at $l_d$ but subdominant to the instantaneous ones since $0<\gamma < 1$. Finally, in region III, $h_i$ are irrelevant and renormalized to zero, i.e., $l_p = l_d$ with $\gamma < 0$. Thus, only case I gives the phonon-driven phases. We note that the precise phase boundary will be affected by the initial part of the RG flow, where the coupling constants do not follow the ansatz, Eq.~\eqref{rhos}.

First, let us study the undoped case. The values of $\beta_i$ for the $D$-Mott and CDW phases are 
\begin{equation}
(\beta_1, \dots, \beta_6) =
\begin{cases}
 \left(-\frac{1}{6}, \frac{7}{6}, \frac{7}{6}, -\frac{1}{6},-\frac{1}{6},-\frac{1}{6} \right) & \text{D-Mott}, \\
 \left(-\frac{1}{3}, \frac{1}{3}, \frac{1}{3}, -\frac{1}{3},\frac{7}{3},\frac{1}{3} \right) & \text{CDW}.
 \end{cases}
\end{equation}
These values are extracted from the numerically integrated RG flow by fitting the ansatz in Eq.~\eqref{rhos} in the asymptotic regime. For the armchair nanotubes, the only nonzero initial values are $h_2 (0) = -4 \kappa_1 < 0$ and $h_3 (0) = 8 \kappa_2 > 0$. The fact that $h_3 (0)$ is positive indicates that $h_3$ is subdominant in both phases; either case II or III in Fig.~\ref{Seidel_fig}. In contrast, since $h_2(0)$ is negative, $h_2$ can be dominant depending on the values of $\beta_2$. Indeed in the $D$-Mott phase, $h_2$ is always dominant (case I) since $\beta_2 > 1$. On the other hand, in the CDW phase, $\kappa_1$ needs to be larger than a critical value for $h_2$ to be dominant, since $-1 < \beta_2 < 1$.  Since $\kappa_1$ is responsible for the Peierls distortion, this explains why the CDW phase is more stable than the $D$-Mott phase against the Peierls instability. For zigzag nanotubes, the nonzero initial values are $h_2 (0) = 4 \kappa_4$ and $h_6  (0) = 8 \kappa_3$. Since they are both positive, the retarded coupling are either irrelevant or subdominant to the instantaneous coupling (Fig.~\ref{Seidel_fig}). This is why we do not see any effect of e-ph coupling in zigzag nanotubes.

Now we turn to the doped case. The values of $\beta_i$ for the $d$-wave superconducting state and the CDW are
\begin{equation}
(\beta_1, \beta_3, \beta_5) =
\begin{cases}
\left(\frac{3}{4},  \frac{3}{4},-\frac{1}{4} \right) & \text{$d$SC}, \\
\left(0, 0, 2 \right) & \text{CDW}.
 \end{cases}
\end{equation}
For armchair nanotubes, the initial values that are nonzero are $h_1 (0) = - \kappa_1 < 0$ and $h_3 (0) = 2 \kappa_2 > 0$. In both phases, $h_1$ becomes dominant when $\kappa_1$ is above a critical value since $-1 < \beta_1 < 1$, while $h_3$ is again subdominant. Compared to the undoped case, the absence of umklapp processes reduces the values of $\beta$'s. This explains why the doped case is relatively more stable to Peierls instability than the undoped cases. In this sense, the Peierls instability is not simply competing, but also assisted by the electronic scattering processes. For zigzag nanotubes, the nonzero initial values are $h_1 (0) = \kappa_4$ and $h_5  (0) = 2 \kappa_3$. Since they are both positive, this indicates that the retarded coupling are again irrelevant or subdominant.

\section{Conclusion}
\label{Conclusion}
In this paper, using a weak-coupling renormalization group approach, we have studied the influence of e-ph coupling on low temperature phases of metallic single-wall carbon nanotubes. In armchair nanotubes, we find that the phases induced by short-range electronic correlations (e.g., a $D$-Mott phase, $d$-wave superconductivity, and a charge-density wave) turn into a Peierls insulator by e-ph coupling. We show that the intraband scattering modes (stretching, radial breathing, and transverse optical modes) cause the softening, and in particular, the radial breathing mode or the transverse optical mode is expected to soften first. On the contrary to armchair nanotubes, no Peierls instability is found in zigzag nanotubes. This suggests that  the $D$-Mott phase or CDW is more likely to appear in zigzag nanotubes in experiments. By analyzing the structure of the RG equations, we clarify that the specific forms of e-ph coupling in two types of nanotubes lead to such distinct behavior.

\acknowledgments
We thank Hsiu-Hau Lin for helpful discussions. J.O. acknowledges support from Research Foundation for Opto-Science and Technology. L.M. acknowledges the support from the Deutsche Forschungsgemeinschaft (through SFB 925 and EXC 1074) and from the Landesexzellenzinitiative Hamburg, which is supported by the Joachim Herz Stiftung. W.M.H. especially acknowledges the support from Ministry of Science and Technology, Taiwan through Grant No. MOST 107-2112-M-005-008-MY3.

\appendix
\section{Derivation of effective models}
\label{AppA}
\subsection{Armchair nanotubes}
From Eq.~\eqref{gamma}, the eigenvalues of the hopping Hamiltonian for the armchair nanotubes are $E_{\pm}(n,k) = \pm |\gamma_{n} (k)|$ with
\begin{equation}
|\gamma_n (k)|^2 = 1 + 4 \cos(k a /2) \cos(3 n \theta_1 /2) + 4 \cos^2( ka /2).
\label{gamma_armchair}
\end{equation}
We see that only the $n=0$ component is gapless near $k = \pm 4\pi/ 3a$. Since we are interested in the low energy physics, we keep only the massless mode $n = 0$. In real space, the low-energy effective modes are given by partial Fourier transform along the transverse direction \cite{Lin1998a}
\begin{equation}
c_{m \alpha} (\bm{r}) \simeq \frac{1}{\sqrt{N_z N_{\perp}}} \sum_{k} e^{i  k z}c_{m k 0 \alpha} \equiv \frac{1}{\sqrt{N_{\perp}}} d_{m\alpha} (z).
\end{equation}
For a tube with length $L$, $N_z = L/a$ and $N_{\perp} = 2 N_y$. Substituting this to the original Hamiltonian, we find that the effective model is a two-leg ladder model with a lattice constant $b = a/2$ [Fig~\ref{two-leg}(a)],
\begin{equation}
\begin{split}
H_0 &\simeq -J_0 \sum_{z, \alpha} \Big[ \sum_{\pm}d^{\dagger}_{A\alpha}(z) d_{B\alpha} (z\pm b)  \\
&+ d^{\dagger}_{A\alpha}(z) d_{B\alpha} (z) + \text{h.c.} \Big],\\
H_U &\simeq \frac{U}{N_{\perp}} \sum_{z, m}  n_{m\up}(z) n_{m\down} (z), \\
H_V &\simeq \frac{1}{N_{\perp}} \sum_{z, \alpha, \beta} \Big[ V_{\perp}n_{A\alpha}(z) n_{B\beta} (z) \\
&+ V\sum_{\pm} n_{A\alpha}(z) n_{B\beta} (z \pm b) \Big].
\end{split}
\end{equation}
The lattice coordinate is given by $z = p b$ with $p \in \mathbb{Z}$. A more natural labeling is given by using the chain index (1 or 2)
\begin{equation}
\begin{split}
d_{A \alpha} (z) &= 
\begin{cases}
d_{1 \alpha}(z) & (z \in 2 \mathbb{Z}) \\
d_{2 \alpha}(z) & (z \in 2 \mathbb{Z} + 1)
\end{cases},\\
d_{B \alpha} (z) &= 
\begin{cases}
d_{2 \alpha}(z) & (z \in 2 \mathbb{Z}) \\
d_{1 \alpha}(z) & (z \in 2 \mathbb{Z} + 1)
\end{cases}.
\end{split}
\label{relabel}
\end{equation}
Eigenstates of the hopping Hamiltonian are given by $d_{\pm k \alpha} = (d_{1k\alpha} \pm d_{2k\alpha})/\sqrt{2}$, and the eigenvalues are $E_{\pm} (k) = - J_0 \cos(k b) \mp J_0$. We note that $E_- (k)$ is shifted by $\pi/b$ compared to the one in Eq.~\eqref{gamma_armchair}. This is due to the relabeling of the sites, Eq.~\eqref{relabel}, which leads to
\begin{equation}
\begin{pmatrix}
d_{+ k \alpha}\\
d_{-, k+\pi, \alpha}
\end{pmatrix}
 = \frac{1}{\sqrt{2}} 
 \begin{pmatrix}
 1 & 1 \\
 1 & -1
 \end{pmatrix}
 \begin{pmatrix}
d_{A k \alpha}\\
d_{B k \alpha}
\end{pmatrix}
\label{diag1}
\end{equation}
in momentum space.

\subsection{Zigzag nanotubes}
For the zigzag nanotubes, we have 
\begin{equation}
|\gamma_n (k)|^2 = 1 + 4 \cos^2(n \theta_z/2) + 4 \cos(n \theta_z/2) \cos(\sqrt{3} k a /2),
\end{equation}
from Eq.~\eqref{gamma}. The condition $\cos(n \theta_z/2) = \pm 1/2$ indicates that the gapless mode appears when $N_x$ is divisible by three. Here we focus on the gapless modes with $n_0 \equiv \pm N_x/3$. The low-energy modes are approximated as 
\begin{equation}
\begin{split}
c_{m \alpha} (\bm{r}) &\simeq \frac{1}{\sqrt{N_z N_{\perp}}} \sum_{k, n=\pm n_0} e^{i  (k z + n \theta)}c_{m k n \alpha}  \\
&\equiv \frac{1}{\sqrt{N_\perp}} \sum_{q = \pm} e^{ i q n_0\theta } d_{m q \alpha} (z).
\end{split}
\end{equation}
For a tube with length $L$, $N_z = L/\sqrt{3}a$ and $N_{\perp} = 2 N_x$. Substituting this to the original Hamiltonian, we find\cite{Bunder2008}
\begin{equation}
\begin{split}
H_0 &\simeq -J_0 \sum_{z, q, \alpha} \Big[ d^{\dagger}_{Aq\alpha}(z) d_{Bq\alpha} (z-b'_+) \\
&+ d^{\dagger}_{A q \alpha}(z) d_{B q \alpha} (z+b'_-) + \text{h.c.} \Big],\\
H_U &\simeq \frac{U}{N_{\perp}} \sum_{z, q, m} \Big[ n_{mq\up}(z) n_{mq\down} (z) +n_{mq\up}(z) n_{m\bar{q}\down} (z)\\
&+ d_{mq\up}^{\dagger} (z) d_{m\bar{q}\up} (z) d_{m\bar{q}\down}^{\dagger} (z) d_{mq \down} (z) \Big],\\
H_V &\simeq \frac{V}{N_{\perp}} \sum_{z, q, q', \alpha, \beta} \Big[ 2n_{A q\alpha}(z) n_{Bq'\beta} (z + b'_-) \\
&-\delta_{q',\bar{q}} d_{Aq\alpha}^{\dagger} (z) d_{Aq'\alpha} (z) d_{Bq'\beta}^{\dagger} (z+b'_-) d_{Bq \beta} (z+b'_-) \Big] \\
&+ \frac{V_{\perp}}{N_{\perp}}  \sum_{z, q,q', \alpha, \beta} \Big[ n_{A q\alpha}(z) n_{Bq'\beta} (z - b'_+) \\
&+ \delta_{q',\bar{q}}  d_{Aq\alpha}^{\dagger} (z) d_{Aq'\alpha} (z) d_{Bq' \beta}^{\dagger} (z-b'_+) d_{Bq \beta} (z-b'_+) \Big],
\end{split}
\end{equation}
where $\bar{q} = -q$, $b' = a\sqrt{3}/4$, $b'_{\pm} = b' \pm \delta$, and $\delta = a/4 \sqrt{3}$. The lattice coordinate is given by $z = 2 p b'$ with $p \in \mathbb{Z}$; the new lattice constant is $2b'$ [Fig.~\ref{two-leg}(b)]. In zigzag cases, the two species $q=\pm$ are decoupled in $H_0$. For each species, eigenstates of the hopping Hamiltonian are then given by 
\begin{equation}
\begin{pmatrix}
d_{R q k \alpha}\\
d_{L q k \alpha}
\end{pmatrix}
 = \frac{1}{\sqrt{2}} 
 \begin{pmatrix}
 1 & e^{-i k \delta} \\
 1 & -e^{-i k \delta}
 \end{pmatrix}
 \begin{pmatrix}
d_{A q k \alpha}\\
d_{B q k \alpha}
\end{pmatrix}
\label{diag2}
\end{equation}
and the eigenvalues are $E_{R/L} = \mp 2 J_0 \cos(kb')$. The Fermi point is at or near $\pi/2b'$, i.e., the Brillouin zone boundary.

\subsection{Effective theory}
In order to obtain the low energy effective theory, we first take the continuum limit, 
\begin{align}
b^{(\prime)} &\sum_z \rightarrow \int dz, & \frac{1}{\sqrt{b^{(\prime)}}} d (z) \rightarrow \psi (z).
\end{align}
$b^{(\prime)}$ is a effective lattice constant for gapless modes. Then we linearize the dispersion around the Fermi energy, and  introduce two chiral fields $\psi_{R, L} (z)$, which are slowly modulating compared to $1/k_F$. For armchair nanotubes, we have 
\begin{equation}
d_{q \alpha} (z)/\sqrt{b} \sim \psi_{Rq\alpha} (z)e^{ik_F z} +\psi_{Lq \alpha} (z) e^{-i k_F z}.
\label{chiral1}
\end{equation}
For zigzag nanotubes, we have \cite{Lin1998a, Bunder2007}
\begin{equation}
\begin{split}
&d_{qA\alpha} (z)/\sqrt{b'} \sim \psi_{Rq\alpha} (z)e^{ik_F z} +\psi_{L\bar{q} \alpha} (z) e^{-i k_F z}, \\
&d_{qB\alpha} (z\pm b' - \delta)/\sqrt{b'} \\
&\sim \psi_{Rq\alpha} (z\pm b')e^{ik_F (z \pm b')} +\psi_{L\bar{q} \alpha} (z \pm b') e^{i k_F (z \pm b')},
\end{split}
\label{chiral2}
\end{equation}
where the factor $e^{-i k \delta}$ in the Fourier transform is canceled due to Eq.~\eqref{diag2}. We note that we use a special labeling of left-moving fields, $d_{qL\alpha}(z) \rightarrow \psi_{\bar{q}L\alpha}$, for zigzag cases. This is more convenient since we can treat armchair and zigzag nanotubes in the same manner. The final results do not change even if we use more natural labeling, $d_{qL\alpha}(z) \rightarrow \psi_{qL\alpha}$. The kinetic term now looks
\begin{equation}
H_0 = v \sum_{q, \alpha} \int dz  \left(\psi_{R q \alpha}^{\dagger} i \partial_z \psi_{R q \alpha} - \psi_{L q \alpha}^{\dagger} i \partial_z \psi_{L q \alpha} \right),
\end{equation}
where $v$ is the Fermi velocity. Finally, substituting the chiral decompositions in Eqs.~\eqref{chiral1} and \eqref{chiral2} into the interaction Hamiltonian, we obtain Eq.~\eqref{Hint}.


\section{Renormalization group equations}
\label{AppB}
The RG equations can be derived via operator product expansions. \cite{cardy1996scaling, Balents1996, Delft1998, Chang2005} The initial values of the coupling constants for the armchair nanotubes are
\begin{equation}
\begin{aligned}
c_{11}^l &= \Lambda^{-1} (U - V + V_{\perp}), &  c_{11}^s &= \Lambda^{-1} (U + 2V + V_{\perp}), \\
c_{12}^l &= \Lambda^{-1} (U - 2V - V_{\perp}), &  c_{12}^s &= \Lambda^{-1} (U + V - V_{\perp}), \\ 
f_{12}^l &= \Lambda^{-1} (U - 2V - V_{\perp}), &  f_{12}^s &= \Lambda^{-1} (U + 2V + V_{\perp}), \\
u_{11}^+ &= \Lambda^{-1} (U -  2V - V_{\perp}), & u_{12}^+ &= \Lambda^{-1} (U - V + V_{\perp}), \\
u_{12}^- &= \Lambda^{-1} (-U + 2V + V_{\perp}), & &
\end{aligned}
\end{equation}
where $\Lambda = 4\pi v N_{\perp}$. We note that initial values for the armchair and zigzag cases can be mapped to each other via the following transformation,
\begin{equation}
V  \leftrightarrow \frac{1}{3} (V + 2 V_{\perp}), \  V_{\perp} \leftrightarrow \frac{1}{3} (4V - V_{\perp}).
\label{mapping}
\end{equation}

The RG equations for instantaneous coupling constants are 
\begin{equation}\label{2sRGeq}
\begin{split}
\dot{c}^l_{11}&=-2c^l_{12}c^s_{12}-2\left(c^l_{11}\right)^2-2\left(u_{12}^+\right)^2- 2u_{12}^+u_{12}^-,\\
\dot{c}^s_{11}&=-\left(c^l_{12}\right)^2-\left(c^s_{12}\right)^2-\left(c^l_{11}\right)^2+\left(u_{12}^-\right)^2,\\
\dot{c}^l_{12}&=-2c^l_{11}c^s_{12}-2c^l_{12}c^s_{11}-4f_{12}^lc_{12}^l+2f_{12}^lc_{12}^s\\
&+2f_{12}^s c_{12}^l + 2u_{11}^+u_{12}^-+2u_{11}^+u_{12}^+,\\
\dot{c}^s_{12}&=-2c^l_{11}c^l_{12}-2c^s_{11}c^s_{12}+2f_{12}^s c_{12}^s+2u_{11}^+u_{12}^+,\\
\dot{f}^l_{12}&=-2\left(f_{12}^l\right)^2-2\left(c_{12}^l\right)^2+2c_{12}^lc_{12}^s\\
&-2\left(u_{12}^-\right)^2-2u_{12}^+u_{12}^-,\\
\dot{f}^s_{12}&=\left(c_{12}^s\right)^2-\left(f_{12}^l\right)^2+\left(u_{11}^+\right)^2+\left(u_{12}^+\right)^2,\\
\dot{u}^+_{11}& =4u_{11}^+f_{12}^s-2u_{11}^+f_{12}^l+2u_{12}^+c_{12}^s\\
&+2u_{12}^+c_{12}^l-2u_{12}^-c_{12}^s+4u_{12}^-c_{12}^l,\\
\dot{u}^+_{12}&=2u_{12}^+f_{12}^{s}+2u_{11}^+c_{12}^{s}+2u^+_{12}c_{11}^s\\
&-4u_{12}^+c_{11}^l-2u_{12}^-c_{11}^l,\\
\dot{u}^-_{12}&=-2u_{12}^+f_{12}^{l}-4u_{12}^-f_{12}^{l}+2u_{12}^-f_{12}^{s}\\
&-2u_{11}^+c_{12}^s+2u_{11}^+c_{12}^l+2u_{12}^-c_{11}^s.
\end{split}
\end{equation}

For the retarded coupling constants, we have
\begin{equation}\label{2rRGeq}
\begin{split}
\dot{\tilde{c}}^l_{11}&= 2\tilde{c}^l_{11}c^s_{11}-4\tilde{c}^l_{11}c^l_{11}-2\left(\tilde{c}^l_{11}\right)^2-4u_{12}^+\tilde{u}_{12}^+\\
&-2\left(\tilde{u}_{12}^+\right)^2 - 2 \tilde{u}_{12}^+u_{12}^-,\\
\dot{\tilde{c}}^l_{12}&=-4\tilde{f}^l_{12}c^l_{12}-4f^l_{12}\tilde{c}^l_{12}-4\tilde{f}^l_{12}\tilde{c}^l_{12}+2\tilde{f}^l_{12}c^s_{12}+2f^s_{12}\tilde{c}^l_{12} \\
&+2\tilde{u}_{11}^+u_{12}^++4\tilde{u}_{11}^+u_{12}^-+2u_{11}^+\tilde{u}_{12}^-+4\tilde{u}_{11}^+\tilde{u}_{12}^-,\\
\dot{\tilde{f}}^l_{12}&=-4\tilde{f}^l_{12}f^l_{12}-2\left(\tilde{f}^l_{12}\right)^2-4\tilde{c}^l_{12}c^l_{12}-2\left(\tilde{c}^l_{12}\right)^2\\
&+2\tilde{f}^l_{12}f^s_{12}+2\tilde{c}^l_{12}c^s_{12} -4u_{12}^-\tilde{u}_{12}^--2\left(\tilde{u}_{12}^-\right)^2\\
&-2u_{12}^+\tilde{u}_{12}^--2u_{11}^+\tilde{u}_{11}^+-2\left(\tilde{u}_{11}^+\right)^2,\\
\dot{\tilde{u}}^+_{11}&= 2\tilde{u}_{11}^+f^s_{12}-4\tilde{u}_{11}^+f^l_{12}-2u_{11}^+\tilde{f}^l_{12}-4\tilde{u}_{11}^+\tilde{f}^l_{12}\\
&+2u^+_{12}\tilde{c}^l_{12}-2\tilde{u}^-_{12}c^s_{12}+4\left( u^-_{12}\tilde{c}^l_{12}+\tilde{u}^-_{12}c^l_{12}+\tilde{u}^-_{12}\tilde{c}^l_{12}\right),\\
\dot{\tilde{u}}^+_{12}&= 2\tilde{u}^+_{12}c^s_{11}-4\left( \tilde{u}^+_{12}c^l_{11}+u^+_{12}\tilde{c}^l_{11}+\tilde{u}^+_{12}\tilde{c}^l_{11}\right)-2u^-_{12}\tilde{c}^l_{11},\\
\dot{\tilde{u}}^-_{12}&= -2u^+_{12}\tilde{f}_{12}^{l}+2\tilde{u}^-_{12}f_{12}^{s}-4\left( u^-_{12}\tilde{f}^{l}_{12}+\tilde{u}^-_{12}f^{l}_{12}+\tilde{u}^-_{12}\tilde{f}^{l}_{12}\right) \\ 
&-2\tilde{u}^+_{11}c^s_{12}+2u^+_{11}\tilde{c}^l_{12}+4\tilde{u}^+_{11}c^l_{12}+4\tilde{u}^+_{11}\tilde{c}^l_{12}.
\end{split}
\end{equation}

\section{Bosonization}
\label{AppC}
In this section, we summarize the bosonization formula used in the main text. The details of derivations are refereed to Refs.~\onlinecite{giamarchi2004quantum, gogolin2004bosonization, Delft1998, Haldane1981, Voit1995, Carpentier2006}. The bosonic fields are introduced as
\begin{equation}
\psi_{r q \alpha} = \frac{\eta_{q \alpha}}{\sqrt{2\pi \alpha_0}} e^{- i r \Phi_{r q \alpha}},
\end{equation}
where $r = R/L = +/-$ is chirality, and $\alpha_0$ is a cutoff of the order of the lattice constant. The bosonic fields satisfy commutations relations
\begin{equation}
\begin{split}
[\Phi_{r q \alpha} (z), \Phi_{r q' \beta}(z') ] &= i r \pi \delta_{qq'} \delta_{\alpha \beta} \text{sgn} (z-z'), \\
[\Phi_{R q \alpha} (z), \Phi_{L q' \beta}(z') ] &= i \pi \delta_{qq'}\delta_{\alpha \beta}.
\end{split}
\end{equation}
The Majorana fermions take care of the anti-commutative properties of fermions,
\begin{equation}
\{\eta_{q \alpha}, \eta_{q' \beta} \} = 2 \delta_{qq'}\delta_{\alpha \beta}.
\end{equation}
More convenient representation is given by the nonchiral fields
\begin{equation}
\phi_{q\alpha}, \ \theta_{q\alpha} = (\Phi_{Lq\alpha} - \Phi_{Rq\alpha})/2.
\end{equation}
They satisfy 
\begin{equation}
\begin{split}
[\phi_{q\alpha} (z), \phi_{q' \beta} (z')] &=[\theta_{q\alpha} (z), \theta_{q' \beta} (z')] = 0, \\
[\phi_{q\alpha} (z), \theta_{q' \beta} (z')] &= i \pi \delta_{qq'} \delta_{\alpha \beta} \Theta(z'-z),
\end{split}
\end{equation}
where $\Theta(z)$ is the Heaviside step function. Finally we move to a new basis
\begin{equation}
\begin{pmatrix}
 \phi_{c0}\\ 
 \phi_{c\pi} \\ 
 \phi_{s0} \\ 
\phi_{s\pi}
\end{pmatrix} 
= \frac{1}{2}
\begin{pmatrix}
1 & 1 &1  &1  \\ 
 1& 1 &-1  &-1  \\ 
 1& -1 &1  &-1  \\ 
 1& -1 &-1  &1 
\end{pmatrix} 
\begin{pmatrix}
 \phi_{+\uparrow}\\ 
 \phi_{+\downarrow} \\ 
 \phi_{-\uparrow} \\ 
\phi_{-\downarrow}
\end{pmatrix} ,
\end{equation}
where $(c, s)$ represents charge and spin modes, and $(0, \pi)$ does bonding/antibonding combinations. $\theta$'s are transformed in the same manner. The sign of each coupling constant is determined by products of Majorana fermions (Klein factors), and by commutators between different chirality, $[ \Phi_{R q \alpha}(z), \Phi_{L q' \beta}(x') ] = i\pi \delta_{qq'} \delta_{\alpha \beta}$. We take the eigenvalues of Klein factors composed of two Majorana fermions as
\begin{multline}
i=\eta_{+s}\eta_{-s} = \eta_{+\uparrow}\eta_{+\downarrow} \\
=\eta_{+\uparrow}\eta_{-\downarrow} = \eta_{-\uparrow}\eta_{+\downarrow} = -\eta_{-\uparrow}\eta_{-\downarrow}.
\end{multline}

\end{document}